\documentstyle[12pt]{article}

\newcommand{\be}{\begin{equation}}
\newcommand{\ee}{\end{equation}}

\newcommand{\ba}{\begin{eqnarray}}
\newcommand{\ea}{\end{eqnarray}}

\newcommand{\la}{\lambda}
\newcommand{\al}{\alpha}

\newcommand{\Tr}{\rm Tr}
\newcommand{\tr}{\rm tr}

\begin{document}

\hsize36truepc\vsize51truepc
\hoffset=-.4truein\voffset=-0.5truein
\setlength{\textheight}{8.5 in}

\begin{titlepage}
\begin{center}
\hfill \\
\hfill\\
\hfill {LPTENS-01/08}
\vskip 0.6 in
{\large  Characteristic polynomials of  real symmetric random matrices
}
\vskip .6 in
       \begin{center}
{\bf E. Br\'ezin$^{a)}$}  {\it and} {\bf S. Hikami$^{b)}$}
\end{center}
\vskip 5mm
\begin{center}{$^{a)}$ Laboratoire de Physique Th\'eorique, Ecole Normale
Sup\'erieure}\\ {24 rue Lhomond 75231, Paris Cedex 05, France{\footnote{
 Unit\'e Mixte de Recherche 8549 du Centre National de la
Recherche
Scientifique et de l'\'Ecole Normale Sup\'erieure.
 } }}\\
{$^{b)}$ Department of Pure and Applied Sciences, CREST of JST, }\\
{University of Tokyo,
Meguro-ku, Komaba, Tokyo 153, Japan}\\
E-mail: brezin@physique.ens.fr ; hikami@rishon.c.u-tokyo.ac.jp

\end{center}

 \vskip 0.5 cm
{\bf Abstract}
\end{center}

 It is shown that the correlation functions of the random variables
  $\det(\lambda - X)$, in which $X$ is a real symmetric $ N\times N$ random
matrix,
exhibit  universal local statistics in the large $N$ limit. The
derivation relies  on an exact dual representation of the problem: the
$k$-point functions are expressed  in terms of finite
integrals over (quaternionic)
$k\times k$  matrices.  However the control of the Dyson limit, in which
the distance of the various  parameters $\la$'s  is of the order of the
mean spacing, requires an integration over the symplectic group. It
is shown that a  generalization of the Itzykson-Zuber method holds for
this problem, but contrary to the unitary case,  the semi-classical
result requires a {\it finite} number of corrections to be exact.
  We have also considered the problem of  an external matrix source
  coupled to the random matrix, and obtain  explicit integral
  formulae, which are useful for the analysis of the large $N$ limit.
\vskip 14.5pt
\end{titlepage}
\setlength{\baselineskip}{1.5\baselineskip}


\section{  Introduction }

     The spectrum of eigenvalues of complex Hamiltonians are often
modelled by a
     random matrix theory, in which the random matrices belong to
various ensembles according to the symmetries of the physical problem.
    The most common space-time symmetries  of the Hamiltonian lead to
the consideration of ensembles of  real, complex or quaternionic
     random matrices. In the simplest case one considers
 Gaussian probability distributions. This simple choice is in many
cases  sufficient since it is now understood that the local
statistics of the eigenvalues are universal, i.e. largely independent of
the probability distribution. The most commonly studied Gaussian
ensembles , called
    GOE, GUE and GSE, are  invariant under the orthogonal, unitary or
symplectic groups, respectively, and they all have important applications
\cite{Dyson1,Dyson2,Mehta}.

In this article we follow our previous study for the GUE case of the
      characteristic polynomials of random matrices \cite{BH1a}. If
$X$ is an
$N\times N$ random matrix , whose characteristic polynomial is
    $ \det(\lambda - X)$,
 we consider   the average of  products of such characteristic
polynomials
      defined by
     \be  F_{k}(\lambda_1, ...,\lambda_{k}) = <  \prod_{l=1}^{k}
    \det (\lambda_l - X ) > .\ee

 In the GUE case we have derived in a previous article explicit formulae
for those correlation functions, found then their asymptotic behavior for
large N, and proved  their universality  in the short distance limit
in which the differences $\lambda_i-\lambda_j$  is the order of the
mean spacing of the eigenvalues of $X$. As usual the orthogonal and
symplectic cases are more difficult to handle. It turns out that there is
a hidden duality in these problems between $N$ the size of the matrices,
and $k$ the number of points in the correlation functions : we may turn
the integrals over $N\times N$ matrices into integrals over $k \times k$
matrices and, since we are interested in large $N$ -finite $k$ limit, this
is the required tool for obtaining the large $N$ limit by the saddle-point
method.  We return below to the GUE case and exhibit its self-duality.
However the GOE case turns out to be dual to the GSE. This duality, in
the simpler case of all $\lambda_i$'s equal, has been discussed recently
within the orthogonal
polynomial method \cite{MN,BF}. In both cases one
may use  geometry  to reduce further the number of integrations. In the
GUE case it relies  on the Harish-Chandra-Itzykson-Zuber formula (HIZ)
\cite{Harish-Chandra,Itzykson-Zuber}. For the GOE case it turns out
that the $N-k$ duality maps the problem into the GSE case, and the use of
the geometry of the symmetric space  $U(2k)/Sp(k)$ leads to considerable
simplifications. At the orders that we  have considered one finds that
there is a generalization of (HIZ) ; it is well-known that in that  case
the WKB approximation happens to be exact. For the GSE problem that one
finds here, it is WKB, plus a finite number of corrections, which happens
to be exact.
Therefore we shall begin be re-exposing the unitary case at the light of
this duality and of the HIZ formula. We shall then proceed to the GOE
ensemble.
\section {Survey of the unitary ensemble}
For the Gaussian unitary
ensemble (GUE), the random matrix
$X$ is a
     complex $N \times N$ Hermitian matrix, with a probability distribution
function

      \be
       P(X) =\frac{1}{Z} \exp( - {N\over{2}} \tr X^2 ) .\ee
    The average $<...>$ means integration with the normalized weight
$P(X)$ ,  with the Euclidean measure
    $\prod_idX_{ii}\prod_{i<j} d\Re X_{ij}d\Im X_{ij}$.
     It is easily shown that the $F_{2}(\lambda_1,\lambda_2)$ reduces, up
to a trivial
factor, to the
 the kernel \cite{Mehta}
       $K_N(\lambda_1,\lambda_2)$ which characterizes the correlation
functions of the eigenvalues of $X$.
    When all the $\lambda_j$'s are nearby , i.e. in the short distance
    scaling region in which $N$ is large        and the products
$N(\lambda_i-\lambda_j)$ are finite,
    $F_{2k}(\lambda_1,...,\lambda_{2k})$ becomes, within an appropriate
    scaling, a universal function, i.e. independent of the specific
distribution $P(X)$.
When all the $\lambda_j$ are equal,
    \be
      F_{2k}(\lambda) = F_{2k}(\lambda,....,\lambda) = < [\det(\lambda -
X)]^{2k} >
    \ee
    the 2k-th moment of the characteristic polynomial.
    In the large N limit, we have derived earlier \cite{BH1a,BH1b}
    \be \label{5}
        F_{2k}(\lambda) = \gamma_k  [ 2 \pi N \rho(\lambda) ]^{k^2},
    \ee
   in which $\rho(\lambda)$ is the density of eigenvalues, and  $\gamma_k$
is a universal factor. This number had been
first computed
    for the circular unitary ensemble  by Keating and Snaith who used the
Selberg integral
    formula \cite{Keating}.

    There are several different derivations for those results. Let us
here expose the duality which was mentioned in the introduction.
We introduce Grassmann variables $c$ and $\bar c$, normalized to
\be \int dc d\bar c e^{iN \bar c c} = 1 .\ee
Then the characteristic
       polynomial may be written as
      \be \label{det}
        \det (\lambda - X) =
\int \prod_{a=1}^N d \bar c_a d c_a e^{i N \bar
       c_a ( \lambda \delta_{ab} - X_{ab} ) c_b}.
      \ee
 Repeating this k-times
\be \label{Grass}
       \prod_{\alpha =1}^k \det (\lambda_{\alpha} - X) =
\int \prod_{a=1}^N \prod_{\alpha =1}^k d \bar c_{a \alpha} d c_{a\alpha}
e^{i N
      \sum_{a=1}^N\sum_{\alpha =1}^k \bar c_{a\alpha} ( \lambda_{\alpha}
\delta_{ab} - X_{ab} ) c_{b\alpha}}.
      \ee
      The (normalized) integration over $X$ , in
presence of a matrix source
$Y$,
yields
       \be
         \int dX e^{-{N\over{2}} \Tr X^2 + i N \tr X Y} = e^{-{N\over{2}}
        \Tr Y^2 }\ .
       \ee
       We may now apply this  to the matrix $Y_{ab} = -
        \sum_{\alpha=1}^k
        \bar c_{a\alpha} c_{b \alpha}$, generated by (\ref{Grass}).
Then  one finds easily  that
\be
         \Tr ( Y^2 ) = -\sum_{\alpha \beta =1}^k \sum _{a=1}^N \bar
c_{a\alpha} c_{a\beta} \sum _{b=1}^N \bar c_{b \beta} c_{b\alpha} = -
\tr (\gamma^2 )
       \ee
  with
        \be
         \gamma_{\alpha \beta} = \sum_{a=1}^N \bar c_{a\alpha} c_{a
\beta}\ .
        \ee
Our notation are as follows :
"$\Tr$" refers here to
         N-dimensional space, whereas
       "$\tr$" refers to matrices acting in the k-dimensional space.
We introduce next
    an auxiliary matrix $k\times k$ hermitian matrix $B$, such that
\be \label{B}e^{{N\over{2}}
        \tr \gamma^2 } =\int dB \exp{(- \frac{N}{2}\tr B^2 + N \tr
\gamma B)},\ee
integrate over the Grassmann variables (which are now decoupled  in the
original N-dimensional space) and end up with
\ba \label{duality}
F_k(\la_1\cdots \la_k) &=& \int dB \det (\Lambda - iB)^N
\exp{-( \frac{N}{2}\tr B^2)}\\ \nonumber &=& e^{ \frac{N}{2}\tr
\Lambda^2}\int dB
(\det B)^N
\exp{-( \frac{N}{2} \tr B^2 +iN \tr \Lambda B) }\ea
in which $\Lambda$ is the diagonal matrix $(\la_1, \cdots, \la_k)$. The
problem is thus mapped into  Gaussian integrals over
$k\times k$ hermitian matrices as announced.
This dual representation is of course well adapted to the
$k$-fixed, $N$-large, limit that we are considering, since
(\ref{duality}) contains
$k^2$ variables instead of $N^2$ in our starting point. It is not
difficult to proceed from (\ref{duality}) and derive the scaling results
which were given in \cite{BH1a}.

However it turns out that it is
simpler , and necessary in view of what comes next for the GOE problem,
 to integrate out the
unitary degrees of freedom in (\ref{duality}). This is done through the
HIZ formula \cite{Harish-Chandra,Itzykson-Zuber} which gives the
integral over the unitary group $U(k)$ :
\be \label{IZ} \int dU e^{iN \tr UXU^{\dagger} Y} = C_N \frac{\det_{1\leq
i,j\leq k} e^{iN x_iy_j}}{\Delta(x_1, \cdots, x_k) \Delta(y_1, \cdots,
y_k)}
\ee in  which the $x_i$'s and $y_j$'s are the eigenvalues of the
Hermitian
$X$ and
$Y$ respectively; $\Delta(x_1, \cdots, x_k) $ is the Van der Monde
determinant
\be \Delta(x_1, \cdots, x_k) = \prod_{i<j} (x_i-x_j). \ee
It is well-known that the  formula (\ref{IZ}) happens to be exact
semi-classically, i.e. if one retains only the sum of the $k!$ stationary
points in the space of unitary matrices, weighted by the Gaussian
fluctuations around each of them. Higher corrections happen to cancel
exactly.     This  leads immediately to an integral over the
$k$ eigenvalues
$b_l$ of
$B$, rather than over the $k^2$ matrix elements :
    \be
      F_{k}(\lambda_1, \cdots,\lambda_k) = C e^{ \frac{N}{2}\tr
\Lambda^2} \int \left(\prod_{l=1}^{k}db_l b_l^N\right)
       e^{- {N\over{2}} \sum_{l=1}^{k}\left( b_l^2 -2i
b_l\lambda_l\right)}
       \prod_{l<l'}^{k}\frac{(b_l - b_{l'})}{(\lambda_l-\lambda_{l'})}
    \ee
When we consider simply the $k$-th moment of the characteristic
polynomials, namely the case in which ($\lambda = \lambda_1 = \cdots =
\lambda_k$), the previous formula reduces to
\be
     F_{k}(\lambda)= F_{k}(\lambda, \cdots,\lambda) =
C e^{ \frac{Nk}{2}
\lambda^2} \int \left(\prod_{l=1}^{k}db_l b_l^N\right)
       e^{- {N\over{2}} \sum_{l=1}^{k}\left( b_l^2 -2i
b_l\lambda\right)}
       \prod_{l<l'}^{k}(b_l - b_{l'})^2 \ee
Note that the above representations of the
correlation functions of the characteristic polynomials,
in terms of
integrals over
$k$ variables, are  exact  for any size
$N\times N$ of the random matrices. It is then simple to find the large-N
limit of those functions by saddle-point integration. If we focus to even
values of $k$, and substitute $2k$ to $k$ (the odd case is doable of
course, but it leads to an oscillatory behavior)
       the saddle point equation  for each $b_l$ is
       \be
         b_l^2 - i \lambda_l b_l - 1 = 0
       \ee
       whose roots are
       $b_l^{\pm} = (i \lambda_l \pm \sqrt{4 - \lambda_l^2})/2$.
   In the scaling limit, in which
the
$\lambda_l-\la_{l'}$ are  of order
$1/N$, the eigenvalues
$b_j$ are very close and one must pay attention to the Vandermonde
determinant in the integration measure. Finally the leading saddle-points
correspond to equal numbers of $b_l$
        close to either $b^{+}$ or $b^{-}$, with  $b^{\pm} = (i
\lambda \pm \sqrt{4 - \lambda^2})/2$. There are thus
        $\left(\matrix{2k\cr
                          k\cr}\right)= 2k!/k!k!$ saddle-points of equal
weight.
       The combinatorial factor $\gamma_k$ of (\ref{5}) is then simply
       \ba\label{GUEgamma}
       \gamma_k &=& \left(\matrix{2k\cr
                          k\cr}\right) {(h_k)^2
       \over{h_{2k}}}\nonumber\\
       &=& \prod_{l=0}^{k-1} {l!\over{(k + l)!}}
       \ea
       where we have used
       \ba
          h_k &=& {1\over{(2\pi)^{k/2}}}\int_{-\infty}^{\infty}\prod_1^k d
x_j
           e^{-\frac{1}{2}\sum_{i=1}^k x_i^2}
          \prod_{l<l'}^k (x_l - x_{l'})^2\nonumber\\
           &=& \prod_{l=0}^{k} l!
        \ea
        (This  formula is used when we consider the gaussian
fluctuations near the saddle-point
in which $k$ of the $b_j$'s are near  $b^{+}$
         and the  other half are close to the saddle point $b^{-}$.)

       Finally this representation through an integral over a finite
matrix  matrix $B$, may be generalized
        to the case  of an external matrix  source  $A$  coupled to the
random
        matrix $X$ \cite{BH1c,BH2,BH3}.
        \ba\label{unitaryA}
        &&\int dX \prod_{l=1}^{2k} \det (\lambda_l - X) e^{-{N\over{2}}
        \tr X^2 + N \tr A X}\nonumber\\
        &=& {1\over{\Delta(\lambda)}}\int db_l
        \prod_{i=1}^N\prod_{l=1}^{2k} (a_i - b_l) \prod_{l<l'}^{2k}
        (b_l - b_{l'})e^{-{N\over{2}}\sum b_l^2 + i N \sum \lambda_l b_l}
        \ea

       We shall now transpose these  techniques to
        real symmetric random matrices.

       \section{Real symmetric matrices and characteristic polynomials}

     Again we consider
       \be F_k(\lambda_1,...,\lambda_k) = \int dX e^{-{N\over{2}}
      \tr X^2} \prod_{i=1}^k \det(\lambda_i - X), \ee
        in which $\rm {X}$ is a real symmetric $N \times N$ matrix,
$\rm{X =
      X^{T}}$. It is worth remembering that real symmetric matrices form
the Lie algebra of the symmetric space $U(N)/O(N)$   (the Lie algebra of
$U(N)$ consists of $N^2$ complex hermitian matrices ; the imaginary part
of those matrices are the $N(N-1)/2$ antisymmetric generators of $O(N)$ ;
the real parts are the $N(N+1)/2$ real symmetric generators of the coset)

       Using again Grassmann variables, and the representation (\ref{det})
of the characteristic determinants we are led again to an integration over
real symmetric matrices in the presence of the matrix source
$Y = -
        \sum_{\alpha=1}^k
        \bar c_{a\alpha} c_{b \alpha}$. This gives
       \be
         \int dX e^{-{N\over{2}} \Tr X^2 + i N \tr X Y} = e^{-{N\over{4}}
        \Tr ( Y^2 + Y Y^T)}\ .
       \ee
       We have dealt earlier with
       \be
         \Tr ( Y^2 ) = - \tr ( \gamma^2 )
       \ee
        which led to the integral (\ref{B}) over an hermitian $k\times k$
matrix. In addition we have here
         \be
           \Tr ( Y Y^T ) = \tr ( U V )\ ,
         \ee
         with the  matrices $U$ and $V$  defined by
        \be
           U_{\alpha\beta} = \sum_{a=1}^N \bar c_{a \alpha} \bar c_{a \beta}
        \ee
         \be
             V_{\alpha\beta} = \sum_{a=1}^N c_{a\alpha} c_{a \beta}
         \ee

         Defining the complex conjugation of Grassmann variables  as
          $(\bar c_1 c_2)^{*} = \bar c_2 c_1$, we have $\gamma =
          \gamma^{\dagger}$,
         $V^{\dagger} = U$. Therefore, we may again decompose the remaining
quartic terms in the $c$'s and $\bar c $'s as
        \be
         e^{-{N\over{4}}\tr ( U V )} = \int dD e^{- N \tr (D^{\dagger} D +
         {i\over{2}}
         V^{\dagger} D + {i\over{2}} D^{\dagger} V)}
         \ee
         where $\rm{D}$ is a complex $k \times k$ antisymmetrix matrix,
         $\rm{D = - D^{T}}$.
       Then we have
        \ba\label{BDint}
         F_k(\lambda_1,...,\lambda_k) &=& \int
         \prod d \bar c_{a \alpha} d c_{a \alpha}
            e^{i N \sum_{\alpha = 1}^k \sum_{a = 1}^N \lambda_{\alpha}
            \bar c_{a \alpha} c_{a \alpha}}\nonumber\\
          &\times& \int d B d D e^{-N \tr (B^2 + D^{\dagger} D) + N \sum
            \bar c_{a \alpha}
          c_{a \beta} B_{\beta \alpha}}\nonumber\\
          &\times& e^{-{i\over{2}}D^{\dagger}_{\alpha \beta} c_{a \beta}
            c_{a \alpha}
          -{i\over{2}}D_{\beta \alpha} \bar c_{a \alpha} \bar c_{a \beta}}.
         \ea
       Those auxiliary matrices B and D allow us to integrate over each
pair
$\bar c_a, c_a$ independently of the
other pairs. It is convenient to define
          \be
             \psi_a = \left(\matrix{\bar c_a\cr
                          c_a\cr}\right).
          \ee
         For a given antisymmetric matrix M, ($M = - M^{T}$), we then  have
 the following formula :
          \ba\label{BDint2}
         &&\int \prod d \bar c_{a \alpha} d c_{a \alpha} e^{{iN\over{2}}
          \psi_{\alpha a} M_{\alpha \beta}
          \psi_{\beta a}} \nonumber\\
          &=& [ \int d \bar c_\alpha d c_\alpha
          e^{{iN\over{2}}\psi_\alpha M_{\alpha
          \beta} \psi_{\beta}} ]^N\nonumber\\
          &=& [ - {\rm Pf} M ]^N = [\det M ]^{N/2}
         \ea
          where $\rm{Pf}$ is the pfaffian of the antisymmetrix matrix $M$.
Applying this to our problem,
we deal here with
         \be \label{M}
            M =\left(\matrix{ D &  \Lambda - i B^T\cr
                        -(\Lambda - i B)& D^{\dagger}\cr}\right)
          \ee
         in which  $\Lambda = \rm{diag}(\lambda_1,...\lambda_k)$. Thus we
finally obtain
        \be\label{FBD1}
           F_k(\lambda_1,....,\lambda_k) = \int dB dD e^{-N \tr (B^2 +
           D^{\dagger} D)}[ - {\rm Pf} M]^N
        \ee
           This representation (\ref{FBD1}) of $F_k$ in terms of a finite
number of integrals, here $2k^2-k$
integrals (one $k\times k$ hermitian matrix B, one complex antisymmetric
$k\times k$ matrix D), again exact for any $N$, is a solution to the
problem. However, contrary to  the GUE case, it turns out out that a
direct use of the saddle-point equations fail in the scaling limit. In
other words every term of the perturbative expansion around the
saddle-point turns out to be relevant in the regime in which the products
$N(\la_i-\la_j)$ are finite.

 One may use the remaining invariances of this
representation, to reduce further the number of integrations.  A unitary
transformation $U$, among the $c_a$, which diagonalizes the Hermitian
matrix B, one of the block matrices of $M$,
            transforms   $D$ into  $D
         \rightarrow
          U^{*} D U^{T}$; in other words one can diagonalize $B$ and keep
for $D$
an antisymmetric matrix.  Therefore, applying the
         Harish-Chandra-Itzykson-Zuber formula
\cite{Harish-Chandra,Itzykson-Zuber}
          for the integration over the unitary group, i.e. over the
relative unitary transformation between the
diagonal matrix $\Lambda$ and the eigenbasis of $B$,  we obtain
        \ba\label{differentlambda}
           F_k(\lambda_1,...,\lambda_k) &=& {e^{N \Tr
          \Lambda^2}\over{\Delta(\Lambda)}}
          \int \prod_{l=1}^k d b_l \Delta(b) \int d D e^{-N\sum b_\alpha^2 +
          2 i N \sum \lambda_\alpha b_\alpha - N \Tr D^{\dagger}
           D}\nonumber\\
          &\times & (- {\rm Pf} M )^N
         \ea
           where now the matrix $B$ in $M$ is diagonal ; we have reduced
the integrations to   $k^2$ variables, instead of  $2k^2-k$. However it
turns out that this is still unsufficient : the inapplicability of the
saddle-point method in the scaling limit is still a problem if we proceed
from (\ref{differentlambda}).
It is thus necessary to return to the underlying geometry of the space of
matrices
$M$ in the representation (\ref{M}). In order to make the quaternionic
structure more apparent we return to (\ref{BDint}) and define the spinor
\be \psi_{a\al} = \left(\matrix {\bar c_{a\al} \cr c_{a\al}}\right)\ee
and the adjoint
\be \bar \psi_{a\al} = \left(\matrix {-c_{a\al} \cr
\bar c_{a\al}}\right)\ee
Then the quadratic form in the Grassmann variables of (\ref{BDint}) takes
the form (repeated $N$ times for each index $a$ that we drop)
\be  \sum_{\alpha, \beta =1}^k
            \left(\bar c_{ \alpha}
          c_{ \beta} B_{\beta \alpha}
          -{i\over{2}}D^{\dagger}_{\alpha \beta} c_{ \beta}
            c_{ \alpha}
          -{i\over{2}}D_{\beta \alpha} \bar c_{ \alpha} \bar c_{
\beta} +i\la_{\alpha}\bar c_{\alpha}c_{\alpha}\right) =
\frac{1}{2}
\sum_{\alpha, \beta =1}^k
            \bar \psi_{ \alpha}\left( q_{\alpha, \beta}
+i\la_{\alpha}\delta_{\alpha \beta}\right) \psi_{\beta}
          \ee
in which the $q_{\alpha \beta}$ are quaternionic matrix elements, i.e.
linear combination of the Pauli matrices. The identification in terms of
$2\times 2$ matrices is thus
\be q_{\alpha, \beta} = \left(\matrix  {B_{\alpha, \beta} & -iD_{\alpha,
\beta}^{*}\cr -iD_{\alpha, \beta}& B_{\alpha, \beta}^{*}\cr} \right) .\ee
This defines a self-dual quaternion matrix  \cite{Dyson1,Mehta} , i.e.
\be q_{\alpha \beta}^{\dagger} = q_{ \beta \alpha} \ee
and
$ q_{\alpha \beta} q_{ \beta \alpha}$ is a multiple of identity. Let
$\tilde M_0$ be the quaternionic matrix whose elements are the quaternion
$q_{\alpha \beta}$ and $\tilde M$ the quaternionic matrix with elements.
\be \tilde M_{\alpha \beta} = q_{\alpha \beta} + i
\la_{\alpha}\delta_{\alpha \beta}.\ee
The Grassmannian integration leads
to
\be \int \prod dc_{\alpha}dc_{\alpha} \exp{  -\frac{N}{2} \sum_{\alpha,
\beta =1}^k
            \bar \psi_{ \alpha} \left(q_{\alpha,
\beta}+i\la_{\alpha}\delta_{\alpha \beta}\right)
\psi_{\beta}
          } = Q\det\tilde M = -\rm{Pf} M, \ee
in which $Q \det$ denotes the quaternionic determinant 
 \cite {Dyson1,Mehta}. In addition
\be tr (B^2 + D D^{\dagger}) = \tr (\tilde M_0^2) = \sum_{\alpha \beta}
q_{\alpha \beta} q_{ \beta \alpha}  .\ee

It may be clarifying to show this quaternionic construction explicitely
for the
$k=2$ case. There one has
\be M =\left(\matrix {0&d& \la_1-iB_{11}&-iB_{21}\cr -d&0& -iB_{12}&
\la_2-iB_{22}\cr -\la_1+iB_{11}&iB_{12}&0&-d^{*}\cr iB_{21}& -\la_2
+iB_{22} & d^{*}&0}\right), \ee
and
\be -\rm{Pf} M = \vert d\vert ^2 + (\la_1 -i B_{11})(\la_2-i B_{22}) +
\vert B_{12}\vert^2 .\ee
The equivalent quaternionic construction is
\be \tilde M = \left( \matrix {q_{11} +i\la_1& q_{12}\cr q_{21}&q_{22}
+i\la_2} \right)
\ee
with
\ba q_{11} &=& B_{11} {\mathbf{1}}\  ,\  q_{12} =(\Re B_{12}){\mathbf{1}}
+ i (\Im B_{12}) \sigma_3  + i( \Re d )\sigma_1 +i (\Im d
)\sigma_2\nonumber
\\ q_{21} &=& q_{12}^{\dagger}\  \hspace{ 4mm},\  q_{22} = B_{22}
{\mathbf{1}}
\ea
and
\ba Q\det (\tilde M) = (q_{11} +i \la_1) (q_{22} +i\la_2) - q_{12}
q^{\dagger}_{12} \nonumber \\= (B_{11}+i\la_1)(B_{22}+i\la_2)
+\vert B_{12}\vert^2 + \vert d\vert ^2 = - \rm{Pf} M .\ea  Therefore we
end up for the correlation functions with the following duality :
\be\label{Symp}
         F_k(\lambda_1,...,\lambda_k) = \int
         d\tilde {M}
           (Q\det \tilde{M})^N e^{-N \tr (\tilde {M_0}^2)} . \ee
The original integral over real symmetric $N \times N$ matrices is
replaced by integrals over quaternionic matrices which depend upon
$(2k^2-k)$ degrees of freedom. Those matrices  are the generators for
the symmetric space $U(2k)/Sp(k)$.

This representation (\ref{Symp}) in terms of a   finite number of
integration variables is a priori well adapted to the large N-limit.
However it turns out that in the scaling limit of interest, the
contributions of the non-gaussian fluctuations around the saddle-points
are all relevant. Therefore it is necessary to eliminate first the
"angular" degrees of freedom. When all the  $\la_i$ are equal, namely if
we consider the moments of the characteristic polynomials, one can simply
diagonalize the symplectic matrices in terms of $k$ eigenvalues and then
proceed to the large N limit. This is done in the next section. However
if the $\la_i$'s are unequal we need some equivalent of the HIZ
formalism, which will be described afterwards.
\section{Moments of the characteristic polynomials}

       We first note the trivial $k=1$ case : $F_1(\lambda)= \langle \det
(\la -X)\rangle$ is simply
       \be
         F_1 (\lambda) = \int_{-\infty}^{\infty} db e^{-N b^2} (\lambda - i
         b)^N \ ,
        \ee
      which, up to a trivial factor, is  the Hermite polynomial
${\rm H_N(\sqrt{N}\lambda)}$ which has an oscillatory behavior for large
$N$ when $\la$ belongs to the support of Wigner's semi-circle. Therefore
we consider from now on the more interesting even correlation functions.

      When all the $\la_i$'s are equal,  the matrix $\Lambda$ is a
multiple of identity, and  one can diagonalize the quaternionic matrix
$\tilde M$  through a transformation belonging to the symplectic group
$Sp(2k)$.
        The transformation of $\tilde M$ into the diagonal matrix $T =
\rm{diag}(t_1,...
        t_k)$  yields  the Jacobian $J = [\Delta(t)]^4$,
($\Delta(t)$ is the Vandermonde
        determinant $\prod_{i<j} (t_i-t_j)$).

This gives simply

        \ba\label{F}
           F_{2k}(\lambda) &=& \langle \det (\la -X)^{2k}\rangle \nonumber
\\&=&C
\int
\prod_{l=1}^{2k} d t_l
          (\prod_{l=1}^{2k} t_l )^N \prod_{l<l'}(t_l - t_{l'})^4
           e^{-{N}\sum t_l^2 + i 2 N \lambda \sum t_l }
         \ea

         The integral representation  (\ref{F}) is well suited to the
study of the
         large N limit.
         Exponentiating $t_l^N$ term as $ e^{N {\rm log} t_l}$,
         the integrand is of the form $\displaystyle
\exp({-N\sum_1^{2k}f(t_l)})$ with
         \be
            f(t) =   t^2 - 2 i \lambda  t - \log t
         \ee
          The saddle points for every $t_l$ are solutions of $f'(t_c) =
0$,i.e.
         \be
              2 t^2 - 2 i \lambda t  - 1 = 0
         \ee
         The two solutions are given by
         \be
              t^{\pm} = {1\over{2}} ( i \lambda \pm \sqrt{ 2 - \lambda^2})
          \ee
           The difference  $(t^{+}-t^{-})$ is proportional to the
semi-circular density of eigenvalues of the GOE
ensemble :
          \be
               t^{+} - t^{-} =  \pi \rho(\lambda)
          \ee
            where $ \rho(\lambda) = \sqrt{2 - \lambda^2}/\pi $.
           Expanding $t_l$ around either $t^{+}$ or $t^{-}$, we find that
the leading saddle-points
are those in which  half of the
          $t_l (l = 1,...,2k)$ are near $t^{+}$ and the remaining half near
          $t^{-}$.
          (Other choices give oscillatory contributions in $\displaystyle
\exp({-N\sum_1^{2k}f(t_l)})$ , which damp the
large $N$-limit). Therefore we have to add the
               $ (2k)!/(k!k!)$ leading saddle-points corresponding to the
distribution of half of the $t_l$'s
near $t^+$, and the other half near $t^-$.  The measure term
given by the
           4-th power of the Vandermonde determinant, yields a factor $(\pi
          \rho(\lambda))^{4k^2}$
          from the $k$ variables near $t^{+}$ and the  $k$ near $t^{-}$.
           The exponent $f$ is then expanded around $t^{+}$ or $t^{-}$, and
the
           remaining integral  factorizes into an integration
           around $t^{+}$
           and another one around $t^{-}$.
           The integration around $t^{+}$ is
           \be
             \int_{-\infty}^{\infty}\prod_{l=1}^k d t_l
             e^{- {N\over{2}} f^{\prime\prime}(t^{+}) (t - t^{+})^2}
             \prod_{i<j}^k (t_i - t_j)^4
             = ({1\over{\sqrt{N f^{\prime\prime}(t^{+})}}})^{2 k^2 - k}
                \prod_{l=1}^k (2l)!.
           \ee
            Noting $t^{+}t^{-} = -1$, and $t^{+} - t^{-} = \pi
            \rho(\lambda)$,
             we find $ f^{\prime\prime}(t^{+}) f^{\prime\prime}(t^{-})
               = (\pi \rho(\lambda))^2$.
            We need  to fix the normalization constant C in (\ref{F}).
            It is obtained from
           the integral,
           \be
              \int \prod dt_l e^{-{N\over{2}}\sum t_l^2}\prod_{i<j}^{2k}
            (t_i - t_j)^4
              = ({1\over{N}})^{4 k^2 - k} \prod_{l=1}^{2k} (2 l)!
           \ee
           The constant $C$ in (\ref{F}) is thus the inverse of this number.
           This constant appears as a normalization for the
           n-point correlation function of the Gaussian symplectic
           ensemble \cite{Mehta}.
           Thus including this normalization constant, $F_{2k}(\lambda)$
becomes
            in the large N limit as
           \be
             F_{2k}^{(GOE)} = \gamma_{k} N^{2 k^2} (2 \pi \rho(\lambda))^{2
           k^2 + k}
           \ee
          \ba
              \gamma_{k} &=& {(2k)!\over{k! k!}} {[\prod_{l=1}^k (2
             l)!]^2\over{
            \prod_{l=1}^{2k} (2 l )!}}\nonumber\\
             &=& \prod_{l=1}^k {(2 l - 1)!\over{(2k + 2 l - 1)!}}
            \ea
          For example, in the case of $F_2(\lambda) = < (\det(\lambda -
           X))^2 >$,
           it gives
          \be
              F_2(\lambda)^{(GOE)} = {1\over{6}} N^2 (2 \pi \rho(\lambda))^3
          \ee
            This result agrees with the result which one would deduce from
$ {\rm lim}\
           \rho(\lambda_1,\lambda_2)/
            (\lambda_1 - \lambda_2)$, where the limit means
           $\lambda_1\rightarrow \lambda
          $ and $\lambda_2 \rightarrow \lambda$, ( details are given
         in  appendix A).

           The value of $\gamma_{k}$ agrees with the result of
           COE
           (circular orthogonal ensemble) found
           by Keating and Snaith \cite{Keating} through the Selberg integral
           formula. In the COE  case, however,
           the density of state is a constant, and the factor
           $\rho(\lambda)$
           is absent.
           This result is to be compared with the earlier result for the
GUE,

           \be
             F_{2k}^{(GUE)} = \gamma_{k} (2 N \pi \rho(\lambda))^{k^2}
            \ee
             where $\gamma_{k}$ is given by (\ref{GUEgamma}).
            This universal constant $\gamma_k$ appears also in the average
of the
            moments of the
            Riemann $\zeta$-function, and it has a number theoretical
             meaning
            \cite{BH1a,Keating,ConreyFarmer}.


  \section{Correlations of characteristic  polynomials}

          The integral representation (\ref{duality}) of the
correlations functions
$F_k(\lambda_1,...,\lambda_k)$ is not unitary invariant unless the
$\la_i$'s are all equal. Therefore if we parametrize the matrix $B$ as
$B=U^{\dagger}b U$, in which $b$ is a diagonal matrix, we have to consider
the HIZ integral
\be\label{gint}\int dU \exp{i \tr U^{\dagger}b U \Lambda} = \frac{\det
\exp{i
\la_ib_j}}{\Delta (b) \Delta(\la)}\ee
which is well-known to be WKB exact . This explains why , in the GUE case,
it is equally possible, to apply the saddle-point method with or without
integrating out the unitary group.

In the symplectic case we are not aware of any similar explicit result ;
however it will be shown now, at least for the lowest values of $k$, that
the integral over the symplectic group can be performed exactly. The
result is remarkably that, in this case, WKB plus a {\it{finite}} number
of corrections is exact.

          In the   $k=2$ case,
           we have
          \be\label{defF2}
            F_{2}(\lambda_1,\lambda_2) = < \det(\lambda_1 - X)
           \det(\lambda_2 - X) >
          \ee
           As shown in the previous section, it is given by the integral
(\ref{F}).
           We first evaluate explicitly the angular integral.
           We first diagonalize $B$ by a unitary transformation, and
then write the eigenvalues
             $b_1$ and $b_2$ in terms of new parameters $t$ and $c$,
           \ba
             b_1 &=& (1 - c) t_1 + c t_2 \nonumber\\
             b_2 &=&  c t_1 + (1 - c) t_2
           \ea
           Then we have $b_1 + b_2 = t_1 + t_2$, and $b_1 - b_2 = (1 - 2 c)
           (t_1 - t_2)$. Since the integrand is a function of $\vert
d\vert^2$, we
            change  variable to $|d|^2 = c(1 - c) (t_1 - t_2)^2$ ; then we
have
           $b_1^2 + b_2^2 + 2 |d|^2 = t_1^2 + t_2^2$, and $b_1 b_2 - |d|^2 =
           t_1 t_2$.
           Finally since the parameter $c$ is restricted to the interval
 $0 < c < 1$, we replace it by $c= \sin^2 \theta$.

           This leads to
           \ba
           F_2(\lambda_1,\lambda_2) &=& \int_0^1 dc \int_{-\infty}^{\infty}
dt_1dt_2 (1 - 2 c)
           {(t_1 - t_2)^3\over{(\lambda_1 - \lambda_2)}} (t_1 t_2)^N
\nonumber\\
           &\times& e^{- N (t_1^2 + t_2^2) + 2 i N (t_1 \lambda_1 + t_2
\lambda_2)
           - 2 i N c (t_1 - t_2) (\lambda_1 - \lambda_2)}
           \ea
           The integration over $c$ yields
           \ba\label{sympIZ}
           F_2(\lambda_1,\lambda_2) &=& \int_{-\infty}^{\infty} dt_1 dt_2
(t_1 t_2)^N
           e^{-N(t_1^2 + t_2^2) + 2 i N (t_1 \lambda_1 + t_2 \lambda_2)}
           \nonumber\\
           &\times& [ \frac{1}{N}({t_1 - t_2\over{\lambda_1 -
\lambda_2}})^2 +
           {i\over{N^2}} {t_1 - t_2\over{(\lambda_1 - \lambda_2)^3}}]
           \ea
           When $\lambda_1 = \lambda_2 = \lambda$, it reduces as expected
to (\ref{F}). This formula may be easily checked for finite values of
$N$ , since it reduces to  Gaussian
           integrals over $t_1$ and $t_2$ ;  for instance in the
simplest  case
$N=1$, it gives
$F_2(\lambda_1,\lambda_2) \sim
\lambda_1
           \lambda_2 + 1$, which agrees with the direct calculation  (in
this case the trivial integral over the real axis
$\displaystyle \int dx (\lambda_1-x)(\lambda_2-x) e^{-\frac{1}{2}x^2} $).

However the representation (\ref{sympIZ}), which is exact for any $N$,
makes it clear\\ i) that the large $N$-limit may be found through a
saddle-point integration over $t_1$ and $t_2$ ; this will  be done below.
\\ ii) that in the universal local limit of interest, in which $N$ goes
to infinity, $\lambda_1-\lambda_2$ goes to zero  and $N(\lambda_1-\la_2)$
remains finite,  the large $N$-limit could not have been taken earlier.
If, for instance, we had used the saddle-point method at the level of
(\ref{differentlambda}), we would have missed  the second
term in the bracket of  (\ref {sympIZ}).  If, at the early level of
(\ref{differentlambda}), we had recognized that the regime of interest
requires to expand beyond the Gaussian approximation to the saddle-point,
it would have appeared unexpectedly that the expansion stops after   the
first correction. Therefore (\ref{sympIZ}) could have been obtained by a
semi-classical approximation with a finite number of corrections, here
just one. This is analogous, although not as simple, to the
Harish-Chandra-Itzykson-Zuber formula for the GUE case, which is
semi-classically exact, without any correction term \cite{Duistermaat}.
           The above integration over $c$ is therefore , for $k=2$, the
corresponding HIZ formula  for the symplectic
           group.

 For higher values of $k$ we need a more elaborate
strategy.
The HIZ formula may be easily derived by considering the Laplacian
operator
\cite{Brezin},
\be
L =  - \frac{\partial^2}{\partial X_{ij}^2}.
\ee
Its eigenfunctions are plane waves
\be L e^{ i N\tr \Lambda X} = (N^2\tr \Lambda ^2) e^{ i N\tr \Lambda X}.
\ee One can construct a unitary invariant eigenfunction of $L$, for the
same energy
$N^2\tr \Lambda^2$, by the superposition
\be I = \int dU e^{iN\tr \Lambda U X U^{\dagger}},
\ee
 which is nothing but the HIZ integral. The integral beeing unitary
invariant, it is a function of the $k$ eigenvalues $t_i$
of $X$.  The same considerations hold  for the three
ensembles
$\beta=1,2$ and 4, corresponding to the orthogonal, unitary  and
symplectic ensemble, with
\be
I = \int e^{N\tr \Lambda g X g^{-1}} dg.
\ee
The Laplacian, expressed in terms of a differential operator on the
eigenvalues $t_i$ reads
\be
[ \sum_{i=1}^k \frac{\partial^2}{\partial t_i^2} + \beta \sum_{i=1,(i\ne
j)}^k
\frac{1}{t_i - t_j} \frac{\partial}{\partial t_i}] I =- \epsilon I\ ,
\ee

with the eigenvalue  $\epsilon$
\be
\epsilon = N^2 \sum_{i=1}^k \lambda_i^2
\ee
The $t$-dependent eigenfunctions of this Schr\"{o}dinger operator have a
scalar product given by the measure
\be \langle \varphi_1\vert \varphi_2\rangle = \int dt_1\cdots dt_k
\vert\Delta(t_1\cdots t_k)\vert^{\beta} \ \varphi_1^{*}(t_1\cdots t_k)
\varphi_2(t_1\cdots t_k)
\ee
The measure becomes trivial if one
multiplies the wave function by
$\vert\Delta\vert^{\beta/2}$ . Thus  if one changes $I(t)$ to
\be \psi(t_1\cdots t_k) =
\vert\Delta(t_1\cdots t_k)\vert^{\beta/2} I(t_1\cdots t_k),  \ee one
obtains the
 Hamiltonian,
\be\label{r2}
[ \sum_{i=1}^k \frac{\partial^2}{\partial t_i^2} - \beta (\frac{\beta}{2} -
1) \sum_{i<j}
\frac{1}{(t_i - t_j)^2}]\psi = -\epsilon \psi.
\ee
For $\beta =2$, the solution is again given by plane waves in the $t_i$
and (taking into account the symmetry under permutations of $I$), one
obtains the HIZ formula.

In the
$\beta = 4$ case, the problem is less trivial, but simple for finite
values of $k$. For $k=2$, a solution of this
equation is
\be
  \psi_0 = e^{iN(\lambda_1 t_1 + \lambda_2 t_2)} ( 1 + \frac{2i}{N(t_1 -
t_2)(\lambda_1 - \lambda_2)})
\ee
The symmetry of $I$ under permutation of thed $t_i$'s leads then to the
solution
\be
\psi = e^{iN(\lambda_1 t_1 + \lambda_2 t_2)} ( 1 + \frac{2i}{N(t_1 -
t_2)(\lambda_1 - \lambda_2)}) + e^{iN(\lambda_1 t_2 + \lambda_2 t_1)} ( 1
+
\frac{2i}{N(t_2 - t_1)(\lambda_1 - \lambda_2)})
\ee
Then,  after multiplication
by the Vandermonde factor, we obtain the required symplectic HIZ formula
(for $k=2$),
\be
 I = \frac{1}{(t_1 - t_2)^2 (\lambda_1 - \lambda_2)^2} \psi
 \ee
For  general $k$ ($\beta$ = 4), the solution of (\ref{r2}) is of the form
\be
  \psi_0 = e^{iN(\lambda_1 t_1 + \cdots + \lambda_k t_k)} \chi
\ee
where $\chi$ satifies
\be \label{diff}
[ \sum_{i=1}^k \frac{\partial^2}{\partial t_i^2} + 2 iN\sum_{i=1}^k
\lambda_i
\frac{\partial}{\partial t_i} - \sum_{i<j} \frac{4}{(t_i - t_j)^2}] \chi = 0
\ee
The operator $\displaystyle  \sum_{i=1}^k \frac{\partial^2}
{\partial t_i^2}  - \sum_{i<j} \frac{4}{(t_i - t_j)^2}$ annihilates
the function $\Delta^{-1}(t_1,\cdots, t_k)$ . Consequently the solution
of (\ref{diff}) may be written
\be\label{f} \chi (t_1\cdots t_k) = \frac { f (t_1\cdots t_k) }{
\Delta (t_1\cdots t_k)} \ee
in which $f(t_1 \cdots t_k)$ is a polynomial of degree $k(k-1)/2$ in the
$t_a$'s. Defining
\be\label{tau} \tau_{ij} = N(\la_i-\la_j)(t_i-t_j) \ee
one finds for $k=3$

  \ba\label{I3}
           &&\chi =
          [ 1 - \frac{2}{i}(\frac{1}{\tau_{12}}
           + \frac{1}{\tau_{23}}
           + \frac{1} {\tau_{31}})\nonumber\\
           && -4( \frac{1}{\tau_{12}\tau_{23}} +
\frac{1}{\tau_{23}\tau_{31}}
            + \frac{1}{\tau_{31}\tau_{12}} ) - {12i}
\frac{1}{\tau_{12}\tau_{23}\tau_{31}}] \ .
\ea
 Again,
as for $k=2$, one sees that the successive terms in the r.h.s. of
(\ref{I3}) are of  same order in the limit of interest, and again
they could have been obtained through a finite number of corrections to
a semi-classical calculation. It is remarkable that the series of
$\chi$ stops at the order the inverse of the  Vandermonde ; thus the
symplectic HIZ integral is expressed as the sum of a finite  number
of terms.  The successive coefficients of each term are determined by
the equation (\ref{r2}).

Using this modified HIZ formula for the symplectic case, we obtain for
the  k=3 case,
$F_3(\lambda_1,\lambda_2,\lambda_3)$
which is expressed by
           \ba\label{F3A}
           &&F_3 (\lambda_1,\lambda_2,\lambda_3) =
           \int dt_1dt_2dt_3 e^{-N(t_1^2 + t_2^2 + t_3^2) + 2i N (\lambda_1
t_1 + \lambda_2 t_2
           + \lambda_3 t_3)} (t_1 t_2 t_3)^N
           ({\Delta(t)\over{\Delta(\lambda)}})^2\nonumber\\
           &\times& [ 1 + i ({1\over{N(\lambda_1 - \lambda_2)(t_1 - t_2)}}
           + {1\over{N(\lambda_2 - \lambda_3)(t_2 - t_3)}}
           + {1\over{ N(\lambda_3 - \lambda_1)(t_3 - t_1)}})\nonumber\\
           &- &  {1\over{N^2(\lambda_1 - \lambda_2)(\lambda_2 -
\lambda_3)(t_1 - t_2)
           (t_2 - t_3)}} - {1\over{N^2(\lambda_2 - \lambda_3)(\lambda_3 -
\lambda_1)
           (t_2 - t_3)(t_3 - t_1)}}\nonumber\\
           & -& {1\over{N^2(\lambda_3 - \lambda_1)(\lambda_1 - \lambda_2)
           (t_3 - t_1)(t_1 - t_2)}}
           - i{3\over{2}} {1\over{N^3\Delta(\lambda)\Delta(t)}}]
           \ea
           where $\Delta(t) = (t_1 - t_2) (t_2 - t_3) (t_3 - t_1)$.
(Note that the symmetries  of the integrand allowed us to keep only the
single solution (\ref{I3}), without adding permutations). For low
values of $N$   (i.e. N=1 or 2, one verifies easily this result by
a direct integration over $1\times 1$ or $2\times 2$ matrices.

Higher values of $k$ may be handled in a similar way, but the
combinatorics become quite heavy. For instance in  an appendix the
solution of the case
$k=4$ is given explicitly and, although again it consists of a finite
number of terms, it is quite cumbersome.

 As is now clear , those integral
       representations make it easy to find the scaling  limit (large N,
finite $N(\la_i-\la_j)$). For instance for $k=2$ one finds
the saddle point values of $t_1$ and $t_2$ from (\ref{sympIZ})
in the large N limit,
\be \label{SP}
t_i = \frac{i}{2} \lambda_i + \frac{\epsilon}{2}\sqrt{2 - \lambda_i^2}
\ee
where $i = 1, 2$ and $\epsilon = \pm 1$. We use the parametrization

$\lambda_i = \sqrt{2} \cos \theta_i$. There are a priori four
saddle-points  given by (\ref{SP}), but the two dominant ones are
$t_1 =
\frac{i}{\sqrt{2}}e^{- i
\theta_1
\epsilon}$,
$t_2 = \frac{1}{\sqrt{2}}e^{i \theta_2 \epsilon}$ for $\epsilon = \pm
1$. In the short distance limit, $N$ large and
$N (\theta_1 - \theta_2) = N \theta_{12}$ finite , we
obtain
\be
F_2(\lambda_1,\lambda_2) = \sum_{\epsilon = \pm 1} e^{- 2 i \epsilon N
\theta_{1
2} \sin^2 \theta}
[ \frac{1}{(N\theta_{12})^2} + \frac{i\epsilon}{2 (N\theta_{12})^3
\sin^2\theta}
]
\ee
where $\theta  = ( \theta_1 + \theta_2 )/2$. The semi-circle  density
of states is given by
$\rho(\lambda) = \frac{\sqrt{2}}{\pi}\sin \theta$,
and $\theta_{12}
= -\frac{1}{\sqrt{2}}\frac{\lambda_1 - \lambda_2}{\sin\theta}$.
Thus we obtain  in the  scaling short distant limit,
\be
F_2(\lambda_1,\lambda_2) = C [ \frac{\cos x}{x^2} - \frac{\sin x}{x^3} ]
\ee
where $x = \pi N (\lambda_1 - \lambda_2)\rho(\lambda)$.
It is interesting to note that this function may be expressed as
a half-integer Bessel function, since
$\displaystyle  (\frac{\cos x}{x^2} - \frac{\sin
x}{x^3}) = -\sqrt{\frac{\pi}{2x^3}} J_{3/2} (x)  $. In the  unitary
case, the sine kernel is  similarly   a half integer Bessel
function since
$\displaystyle \frac{\sin x}{x} = \sqrt{\frac{\pi}{2x}} J_{1/2} (x)$.

\section{Extension to an external matrix source}

   In the GUE case, when an external  matrix source $A$ is coupled to an
Hermitian random matrix $X$,
    as we have discussed earlier,  $F_{2k}(\lambda_1,...,\lambda_{2k})$
    is given by (\ref{unitaryA}). The degrees of freedom provided by the
eigenvalues of
$A$ are useful to study a number of new universality classes \cite{BH1c,BHL}.
For instance  by tuning the eigenvalues
   $a_i$ of the external source matrix $A$, we
   can study the problem of a closing gap  in the spectrum of random
hermitian matrices \cite{BH1c}.
    Thus it is  interesting to
   consider this external source problem for
  real symmetric matrices as well .

     One can always assume that the external source matrix $A$ is
diagonal.
     In the method of integration  over Grassmann variables used in
section three, it is simple
to include the
     external matrix $A$ :
     \be
       \int e^{-{N\over{2}} \tr X^2 + N tr A X + i N \tr X Y} dX
     = e^{-{N\over{2}} \tr [(Y - i A)^2 +  (Y - iA) (Y^{T} - i A)}
     \ee
     where $Y = - \sum_{\alpha} \bar c_{a \alpha} c_{b \alpha}$.
     Since $A$ is  diagonal, the term  $\tr A Y$ gives simply the
     extra term ${\rm exp}[iN \sum
     a_n \bar c_{n \alpha} c_{n\alpha}]$ in the integrand  of $F_k$ in
(\ref{BDint}), . Therefore, repeating the
     calculations of section 2, we find that eq.(\ref{BDint2})
     is modified as follows :

          \ba
         &&\int \prod d \bar c_{j \alpha} d c_{j \alpha} e^{{iN\over{2}}
          \psi_{\alpha j} M_{\alpha \beta}^{(j)}
          \psi_{\beta j}} \nonumber\\
          &=& \prod_{j=1}^N \int d \bar c_\alpha d c_\alpha
          e^{{iN\over{2}}\psi_\alpha M_{\alpha
          \beta}^{(j)} \psi_{\beta}} \nonumber\\
          &=& \prod_j[- {\rm Pf} M^{(j)} ] = \prod_j [\det M^{(j)} ]^{/2}
         \ea
          where ${\rm Pf}M^{(a)}$ is the pfaffian of the antisymmetrix
matrix $M^{(a)}$  given by
         \be
            M^{(j)} =\left(\matrix{ D &  \Lambda + a_j {\bf 1} - i B^T\cr
                        -(\Lambda + a_j {\bf 1} - i B)&
D^{\dagger}\cr}\right).
          \ee
         in which  $\Lambda = \rm{diag}(\lambda_1,...\lambda_k)$ and $a_j
{\bf 1} =
         diag (a_j,...,a_j)$. Thus we finally obtain
        \be\label{FBD}
           F_k(\lambda_1,....,\lambda_k) = \int dB dD e^{-N \tr (B^2 +
           D^{\dagger} D)}\prod_{j=1}^N[ - {\rm Pf} M^{(j)}]
        \ee
        This integral can be expressed in terms of a  quarternion matrix
$Q$, which can be
        diagonalized by the symplectic group $Sp(k)$. When all the
$\lambda_i$'s are  equal
        to a single $\lambda$, we get
        \be
          F_k(\lambda,...,\lambda) = \int
          \prod_{l=1}^k \prod_{j=1}^N (t_l - i a_j) \prod_{l<l'} (t_l -
t_{l'})^4
                e^{-N \sum t_l^2 + 2 i N \lambda \sum t_l} \prod_{l=1}^k
dt_l\ .
        \ee
        For the case of different $\lambda_i$'s , this formula is modified
by an extra factor
        as in the previous section.

        As an example of the usefulness of the above representation, we
choose an external
source with only two opposite eigenvalues $\pm c$, with half of the
eigenvalues equal to $+c$ and
the other
half  to $-c$.
This gives a factor $(t_l^2 + c^2)^{N/2} = \exp {\frac{N}{2}\log{
(t^2+c^2)}}$ in the integrand. Expanding it in powers
of $t^2$, the total coefficient of $t^2$ in the exponent vanishes for $c =
1/\sqrt{2}$. Therefore in the large N limit,
        we obtain at this new critical point
        \be\label{intersection}
          < [\det(X)]^k > = \int e^{-N \tr Q^4} dQ
        \ee
        where $Q$ is a $k\times k$ symmetric quaternionic matrix, and the
        the average $<...>$ is evaluated with the distribution in the
presence of
        the external source whose eigenvalues are $\pm c = \pm 1/\sqrt{2}$.

        One could make other choices for the eigenvalues of the external
source matrix $A$, and  obtain thereby
higher multicritical points with terms such as  $\tr Q^n$
 in the exponent,
        in analogy with the GUE case in an external matrix source
\cite{BH3}.

 \section{Summary}

In this article, an exact representation of the k-point functions
$\langle \prod_{a=1}^k \det(\lambda_a-X)\rangle $, averaged over
$N\times N$ real symmetric random matrices,    has been derived
       in terms of an integral  over  quaternionic $k\times k$
matrices, invariant under the unitary symplectic group. This
representation leads to an easy calculation of the moments of the
characteristic polynomials ($\la_1=\cdots=\la_k$).  In the large
$N$-limit one finds
\be
             F_{2k}^{(GOE)} = \prod_{l=1}^k {(2 l - 1)!\over{(2k + 2 l -
1)!}} N^{2 k^2} (2 \pi \rho(\lambda))^{2
           k^2 + k}  \ ,
           \ee
 to be compared to the earlier result for the
GUE,

           \be
             F_{2k}^{(GUE)} = \prod_{l=0}^{k-1} {l!\over{(k + l)!}} (2 N
\pi \rho(\lambda))^{k^2}.
            \ee
For unequal $\la_a$'s, in spite of the fact that the integral
representation involves a finite number of variables, in the large
$N$-limit the
corrections to the saddle-point, in the scaling regime
$N(\la_i-\la_j)$ finite,  are not negligible. A
generalization of the HarishChandra-Itzyson-Zuber formula
is shown to solve the problem. Remarkably this formula is"nearly"
semi-classical, in the sense that it happens that the semi-classical
expansion terminates after a few terms, a number of terms which
increases with $k$ but not with $N$. Then  the saddle-point method may
easily be applied for large $N$, and this leads to explicit
asymptotic formulae
 for the correlation functions of the characteristic polynomials.
Finally this may be generalized to include an external matrix source in
the probability measure.
 Real symmetric random matrices  appear as  models
        of numerous physical time-reversal invariant Hamiltonians. For
instance  the orthogonal matrix model with an external
        source has been investigated as a model of glassy
        behavior \cite{Parisi}.
        The results of the present work for the moments and for the
        correlation functions in an external source may be of interest
for such problems. \\

\newpage
\setcounter{equation}{0}
\renewcommand{\theequation}{A.\arabic
{equation}}
{\bf Appendix A: {The solution for $k=4$}}
\vskip 5mm

From (\ref{diff}) and (\ref{f}), the polynomial $f$ satisfies
\be
\sum_{i=1}^k \frac{\partial^2}{\partial t_i^2} f +
2 \sum_{i=1}^k (\frac{\partial f}{\partial t_i} + i N\lambda_i f)
 (\Delta \frac{\partial}{\partial t_i} \frac{1}{\Delta})
 + 2 i N\sum_{i=1}^k \lambda_i \frac{\partial f}{\partial t_i} = 0
 \ee
The solution of this equation is obtained by a perturbation expansion
in powers of the  $\lambda_i$'s,
but it ends at the level of the Vandermonde $\Delta(\la_1 \cdots\la_4)$.
Using the  notation of (\ref{tau}), $\tau_{ij} = N (t_i - t_j)
(\lambda_i - \lambda_j)$,
 we obtain
\ba
f &=&C[ 1 - \frac{i}{4} ( \tau_{12}  + \tau_{13} + \tau_{14} + \tau_{23} +
\tau_{2
4}
+ \tau_{34})\nonumber\\
&-&\frac{1}{12} ( \tau_{12}\tau_{13} + \tau_{12}\tau_{14} +
\tau_{13}\tau_{14} +
 \tau_{12}\tau_{23}
+ \tau_{23}\tau_{24} + \tau_{12}\tau_{24}\nonumber\\
&+& \tau_{14}\tau_{34} + \tau_{14}\tau_{24} + \tau_{24}
\tau_{34} + \tau_{23}\tau_{34} + \tau_{13}\tau_{34} + \tau_{13}\tau_{23} )
\nonumber\\
&-&\frac{1}{18}(\tau_{12}\tau_{34} + \tau_{13}\tau_{24} +
\tau_{14}\tau_{23})\nonumber\\
&+& \frac{i}{24} (\tau_{12}\tau_{13}\tau_{14} + \tau_{12}\tau_{23}\tau_{24}
+ \tau_{13}\tau_{23}\tau_{34} + \tau_{14}\tau_{24} \tau_{34})\nonumber\\
&+& \frac{i}{36}(\tau_{12}\tau_{13}\tau_{23} + \tau_{12}\tau_{14}\tau_{24}
+ \tau_{13}\tau_{14}\tau_{34} + \tau_{23}\tau_{24}\tau_{34}\nonumber\\
&+& \tau_{14}\tau_{34}\tau_{23} + \tau_{14}\tau_{24}\tau_{23}
+ \tau_{12}\tau_{24}\tau_{34} + \tau_{12}\tau_{23}\tau_{34}\nonumber\\
&+& \tau_{12}\tau_{14}\tau_{23} + \tau_{13}\tau_{14}\tau_{23}
+ \tau_{12}\tau_{13}\tau_{34} + \tau_{12}\tau_{14}\tau_{34}\nonumber\\
&+& \tau_{13}\tau_{34}\tau_{24} + \tau_{13}\tau_{24}\tau_{23}
+ \tau_{14}\tau_{24}\tau_{13} + \tau_{13}\tau_{12}\tau_{24})\nonumber\\
&+& \frac{1}{72} (\tau_{12}\tau_{23}\tau_{34}\tau_{14} +
\tau_{12}\tau_{13}\tau_
{24}\tau_{34}
 + \tau_{13}\tau_{14}\tau_{24}\tau_{23} +
\tau_{12}\tau_{14}\tau_{24}\tau_{34} +
\tau_{12}\tau_{14}\tau
_{24}\tau_{23}
\nonumber\\
&+&  \tau_{12}\tau_{14}\tau_{24}\tau_{13} +
\tau_{12}\tau_{13}\tau_{23}\tau_{34} +
\tau_{12}\tau_{13}\tau_{23}\tau_{14} +
\tau_{12}\tau_{13}\tau_{23}\tau_{24}+\tau_{12}\tau_{24}\tau_{23}\tau_{34}
\nonumber\\
&+&
\tau_{14}\tau_{24}\tau_{23}\tau_{34}
 + \tau_{13}\tau_{24}\tau_{23}\tau_{34} +
\tau_{12}\tau_{14}\tau_{13}\tau_{34}
 + \tau_{14}\tau_{13}\tau_{23}\tau_{34} +
\tau_{14}\tau_{13}\tau_{34}\tau_{24})
\nonumber\\
 &-& \frac{i}{144}(\tau_{12}\tau_{13}\tau_{24}\tau_{23}\tau_{34}
 + \tau_{12}\tau_{14}\tau_{24}\tau_{13}\tau_{23}
 + \tau_{12}\tau_{14}\tau_{24}\tau_{13}\tau_{34} \nonumber\\
&+&
 \tau_{14}\tau_{13}\tau_{24}\tau_{23}\tau_{34}+
 \tau_{12}\tau_{14}\tau_{13}\tau_{34}\tau_{23} +
 \tau_{12}\tau_{14}\tau_{24}\tau_{23}\tau_{34} )\nonumber\\
 &-&\frac{1}{288} \tau_{12}\tau_{13}\tau_{14}\tau_{23}\tau_{24}\tau_{34}].
 \ea
The HIZ integral is obtained by requiring the symmetry under
permutation of the
$t_i$'s
in the final  expression for $I$, thus from (\ref{f}) (and a replacement
of  $\la$ by $2\la$),
 \be
 I =  C \frac{1}{[\Delta(t)\Delta(\lambda)]^3} ( f(t_1,...,t_k) + {\rm perm.
of}
\hskip3mm  f )
 \ee
 where the last term means that one adds the terms in which
one permutes  the
$t_i$'s for fixed $ \lambda$'s .

\end{document}